  \documentclass{gretsi}

   \usepackage[english,french]{babel}   
  \usepackage{times}			
  
\usepackage{amsfonts}
\usepackage{mathtools}
\usepackage{amsthm}
\usepackage{multirow}
\usepackage{subcaption}
\usepackage{xcolor}

\newcommand{\R}{\ensuremath{\mathbb{R}}}
\newcommand{\N}{\ensuremath{\mathbb{N}}}

\newcommand{\bracket}[1]{\left( #1 \right)}
\newcommand{\bracketn}[1]{\left\{#1 \right\}}

\DeclareMathOperator*{\argmin}{argmin}
\DeclareMathOperator*{\argmax}{argmax}

\usepackage{algorithm,algpseudocode}
 
\newtheorem{prop}{Proposition}
\newtheorem{theo}{Theorem}

\newtheorem{rem}{Remark}

\title{Patch-based image Super Resolution using generalized Gaussian mixture model}

\author{\coord{Dang-Phuong-Lan}{Nguyen}{1,2},
        \coord{Jean-Francois}{Aujol}{1},
    \coord{Yannick}{Berthoumieu}{2}} 

\address{\affil{1}{Université de Bordeaux, Bordeaux INP, CNRS, IMB, UMR 5251, F-33400 Talence, France} 
         \affil{2}{Université de Bordeaux, Bordeaux INP, CNRS, IMS, UMR 5218, F-33400 Talence, France, }} 

\email{lan.nguyen@math.u-bordeaux.fr,
jean-francois.aujol@math.u-bordeaux.fr\\
yannick.berthoumieu@ims-bordeaux.fr}

\frenchabstract{Les méthodes de super résolution d'image visent à recréer une image haute résolution à partir d'une basse résolution. La famille d'approches basée sur les patchs a fait l'objet d'une attention et d'un développement considérable. La technique de minimum de l'erreur quadratique moyenne est une méthode de restauration d'images qui utilise un modèle gaussien de probabilités sur les patchs d'images. Cet article propose un algorithme d'apprentissage d'un modèle conjoint de mélange gaussien généralisé (GGMM) à partir des paires de patchs à basse résolution et des patchs correspondants à haute résolution d'une image de référence. À partir de ce modèle GGMM, l'image haute résolution en utilisant la méthode MMSE.  Nos évaluations numériques indiquent que la méthode MMSE-GGMM se comporte très bien par rapport à l'état de l'art.}  

\englishabstract{Single Image Super Resolution (SISR) methods aim to recover the clean images in high resolution from low resolution observations. A family of patch-based approaches have received considerable attention and development. The minimum mean square error (MMSE) method is a powerful image restoration method that uses a probability model on the patches of images. This paper proposes an algorithm to learn a joint generalized Gaussian mixture model (GGMM) from a pair of the low resolution patches and the corresponding high resolution patches from the reference data. We then reconstruct the high resolution image based on the MMSE method. Our numerical evaluations indicate that the MMSE-GGMM method competes with other state of the art methods.}

\begin{document}
\maketitle
\section{Introduction}
Super resolution is the task to reconstruct the estimate $\hat{X}_{HR}$ of a high resolution (HR) image $X_{HR}$ based on a low resolution (LR) observation $X_{LR}$. Observed LR image is generated by an unknown operator $A$
\begin{equation}
    X_{LR}=AX_{HR}+\epsilon.
\end{equation}

In recent years, various patch-based  super resolution image algorithms have been introduced. Zoran and Weiss \cite{EPLL} proposed using the negative log-likelihood function of a GMM as a regularizer of an inverse problem. The estimated HR image $\hat{X}_{HR}$ is computed by solving
\begin{equation} \label{eq:epll}
    \argmin_{X_H} \left\Vert AX_{HR} -X_{LR} \right\Vert^2 -\lambda \sum_{i\in I} \log p\left(X_{HR,i}\right)
\end{equation}
where $p$ is the probability density function of the GMM and $\left(X_{HR,i}\right)_{i\in I}$ are the patches in the HR image. This method is called expected patch log likelihood (EPLL). C. Deledalle et al. \cite{EPLL-GGMM} extended the EPLL method to a new approach that combines the generalized Gaussian mixture model (GGMM) with EPLL. It is called EPLL-GGMM algorithm. They showed that the GGMM gets the distribution of patches better than a GMM and performs better within the EPLL framework. However, EPLL requires knowledge of the operator $A$, which is not the case in some real applications. Therefore, we investigate the alternative approach of P. Sandeep et al. \cite{Sandeep}, which uses a joint GMM of the concatenated vectors of HR and the corresponding LR patches. Each HR patch is estimated from the LR patch by using the Minimum Mean Squared Error (MMSE) estimator. 
To solve the minimization problem of MMSE, the parameters of the joint GMM are required. 
These parameters are usually learned using the expectation–maximization (EM) algorithm. In the case of generalized Gaussian, there exist several methods to learn the parameters, such as Fixed Point (FP) algorithm \cite{FP-MGGD} and Riemannian Averaged Fixed-Point (RA-FP) algorithm \cite{RA-FP-MGGD}.  Following these works, F. Najar et al. \cite{FP-GGMM} developed a fixed-point (FP) algorithm for learning the  parameters of GGMM. However, \cite{FP-GGMM} uses directly the estimated covariance matrix and the shape parameter of GMM without the weight of the mixture. Besides that, the EPLL-GGMM model \cite{EPLL-GGMM} computes the parameters of the GGMM as the GMM. 
In this paper, we propose an algorithm that combines the EM algorithm and the FP estimator to compute the parameters of GGMM. This algorithm is called FP-EM algorithm.

\textbf{Contribution:} This paper provides a method that uses the MMSE estimator for joint GGMM, called MMSE-GGMM. This method  adapts the FP-EM algorithm to learn the joint GGMM. 

This paper is organized as follows.  FP-EM algorithm to learn the GGMM based on the EM algorithm is the major contribution of this paper and is discussed in Section 2. The MMSE estimator for GGMM is detailed in Section 3. Section 4 derives a method to reconstruct the HR image using joint GGMM. Then Section 5 illustrates the success of our method for super resolution of synthetic data as well as material data with unknown operator $A$. 
 
\section{GGMM learning} \label{section learn GGMM}
In this section, we focus on the parameter estimation of the generalized Gaussian mixture model based on the EM algorithm \cite{DLR1977}. The generalized Gaussian mixture model (GGMM) has a probability density function  
\begin{equation} \label{fx}
    P_\theta\left(x\right)=\sum_{k=1}^K w_k f \left(x|\theta_k \right).
\end{equation} 
$f(x|\theta_k)$ is a probability density function of a generalized Gaussian distribution (GGD)
 \begin{equation}
     f\left(x| \theta_k \right) = \frac{C_p\left(\beta_k\right)}{\left|\Sigma_k \right| ^{\frac{1}{2}} } \exp{\left[ -\frac{1}{2} \left[ \left(x-\mu_k\right)^T \Sigma_k^{-1} \left(x-\mu_k\right) \right]^{\beta_k} \right] } \label{eq:MMGGD}
 \end{equation}
where $w_k$ satisfies $\sum_{k=1}^K w_k=1$ and $w_k >0$, for all $k$ and $\theta_k=\bracketn{ \mu_k,\Sigma_k,\beta_k }$ are the parameters of the $k^{th}$  component of GGMM. In detail, $\mu_k\in\R^p$ is the expected value, $\beta_k\in \left(0,+\infty\right)$  is the shape parameter and $\Sigma_k$ is a $p\times p$ a positive semi-definite covariance matrix. The normalizing constant $C_p\left(\beta_k\right)$ is expressed as:
\begin{equation}
    C_p\left(\beta_k\right)=\frac{\Gamma\left(\frac{p}{2} \right) \beta_k} { \pi ^{\frac{p}{2}} \Gamma\left(\frac{p}{2 \beta_k} \right) 2^{\frac{p}{2\beta_k}} }
\end{equation}
where $\Gamma$ is the gamma function.  
\begin{rem}
The Gaussian model is a special case of the generalized Gaussian distribution with the shape parameter $\beta=1$.
\end{rem} 

In the following, we consider the EM algorithm for estimating the parameters of mixture models. Given samples $x_1,...,x_N$ have p-dimensional generalized Gaussian mixture distribution, we minimize the negative log-likelihood function
\begin{equation}
\mathcal L(\mathbf{w},\mathbf{\Theta})=-\frac{1}{N}\sum_{i=1}^N\log\bracket{  \sum_{k=1}^K w_k f\left(  X_i\mid \theta_k\right)},
\end{equation}
where $\mathbf{w}=(w_1,...,w_k), \mathbf{\Theta}=(\theta_1,...,\theta_k).$ Then, the EM Algorithm for GGMM is read as Algorithm \ref{alg_em_mm}. 
\begin{algorithm}[!ht]
\caption{EM Algorithm for Mixture Model}\label{alg_em_mm}
\begin{algorithmic} 
\State Input: $x=(x_1,...,x_N)\in\R^{p\times N}$, initial estimate $\mathbf{w}^{(0)},\mathbf{\Theta}^{(0)}$.
\For {$n=1,2,...$}
\State \textbf{E-Step:} For $k=1,...,K$ and $i=1,\ldots,N$ compute 
        \begin{equation}
            \alpha_{i,k}^{\left(n\right)} = \frac{w_k^{\left(n-1\right)}  f \left(X_i| \theta_k^{\bracket{n-1}} \right)}{\sum_{l=1}^K w_l^{\left(n-1\right)}  f \left(X_i| \theta_l^{\bracket{n-1}} \right)} \label{eq:alpha}
        \end{equation}
\State \textbf{M-Step:} For $k=1,...,K$ compute
    \begin{equation}
        w_k^{\left(n\right) }=\frac{1}{N} \sum_{i=1}^N  \alpha_{i,k}^{\left(n\right)}
    \end{equation}
    \begin{equation}
        \theta_k^{\left(n\right)}=\argmax_{ \theta_k} \sum_{i=1}^N \alpha_{i,k}^{\left(n\right)} \log  f \left(X_i| \theta_k\right). \label{eq:max_logGGMM}
    \end{equation}
\EndFor
\end{algorithmic}
\end{algorithm}

The interesting step of Algorithm \ref{alg_em_mm} is the second step of M-Step which requires the maximization of a function. If the $\alpha_{i,k}$ are equal for all $i=1,...,N$, the optimization problem \eqref{eq:max_logGGMM} has been solved by B. Wang et al. in \cite{convergent GGD}. In this paper, we generalize the algorithm from \cite{convergent GGD} for different weights $\alpha_{i,k}$ in \eqref{eq:max_logGGMM}.  
\begin{prop} \label{prop:FP_GGMM}
Let $f$ be the generalized Gaussian density function \eqref{eq:MMGGD} and $\alpha_i\in\R_{\geq 0},i=1,...,N$. Given the samples $x_1,...,x_N\in\R^p$, if $\theta=\bracketn{\mu,\Sigma,\beta}$ is a solution of
\begin{equation}
    \argmax_{ \theta} {\sum_{i=1}^N \alpha_{i} \log { f \left(x_i| \theta\right)} }, \label{eq:max_logGGMM_pro}
\end{equation}
they should satisfy the following equations: 
\begin{align}
    \mu&= \frac{\sum_{i=1}^N \alpha_{i}  \delta_{i }^{\beta-1} x_{i} }{\sum_{i=1}^N \alpha_{i} \delta_{i}^{\beta-1} }, \label{eq:solution_muk}\\
    \Sigma&=\frac{\sum_{i=1}^N \alpha_i \delta_i^{\beta-1}(x_i-\mu)(x_i-\mu)^T }{\sum_{i=1}^N\alpha_i} \label{eq:solution_Sigmak} 
\end{align} 
   where $\delta_{i}= \left(x_{i}-\mu\right)^T \Sigma^{-1} \left(x_{i}-\mu\right)$, $\rho>0$. The parameter shape $\beta$ can be computed by using Newton-Raphson method \cite{Newton}. 
\end{prop}
 The proof of the previous proposition can be achieved by setting the gradient of the objective function to zero. Therefore, the solution of the maximization problem \eqref{eq:max_logGGMM} can be generated by the FP Algorithm \ref{alg_FP}. The EM algorithm combined with the FP algorithm for GGMM is called FP-EM algorithm. 
\begin{algorithm}[!ht]
\caption{Fixed point (FP) algorithm for \eqref{eq:max_logGGMM} }\label{alg_FP}
\begin{algorithmic} 
\State Input: $x=(x_1,...,x_N)\in\R^{p\times N}$, initial estimate $\mu_k^{(0)},\Sigma_k^{(0)}\beta_k^{(0)}$.
\For {$r=1,2,...$}
\State Update $\mu_k^{(r+1)}$ by \eqref{eq:solution_muk}
\State Update $\Sigma_k^{(r+1)}$ by \eqref{eq:solution_Sigmak} 
\State Update $\beta_k^{(r+1)}$ by Newton-Raphson method
\EndFor
\end{algorithmic}
\end{algorithm} 
\section{SR Reconstruction}\label{sec: SR}
In the following, we introduce a super resolution method using the joint generalized Gaussian mixture model. This method is an extended version of the MMSE-GMM \cite{Sandeep} for the generalized Gaussian case. We aim to reconstruct the unknown HR image $X_{HR}$ from the LR image $X_{LR}$. We assume that we have given a pair of reference images: HR image$\tilde{x}_{HR}$ and LR image $\tilde{x}_{LR}$ with the magnification factor $q$. The super resolution reconstruction includes the three following steps.
\paragraph{A-Learning a joint GGMM}
In this step, we extract the LR patches $\tilde{x}_{LR,i} \in \R^{\tau^2}$ and HR patches $\tilde{x}_{HR,i} \in \R^{q^2\tau^2}$, $q\in\N, q\leq2,i=1,...,N$ from the given images $\tilde{x}_{LR}$ and $\tilde{x}_{HR}$. We approximate the GGMM of the concatenated vector $x_i= \Big(\begin{array}{c}\tilde{x}_{HR,i} \\\tilde{x}_{LR,i} \end{array}\Big) \in\R^p$, $p=(q^2+1)\tau^2$ using the FP-EM algorithm. Then, we obtain the parameters of GGMM: 
$$w=(w_k)_k,\quad \mu=(\mu_k)_k, \quad \Sigma=(\Sigma_k)_k, \quad \beta=(\beta_k)_k
$$
with
$$
\mu_k=\Big(\begin{array}{c}\mu_{H,k}\\\mu_{L,k}\end{array}\Big),\quad \Sigma=\Big(\begin{array}{cc}\Sigma_{H,k}&\Sigma_{HL,k}\\\Sigma_{HL,k}^T&\Sigma_{L,k}\end{array}\Big).
$$ 
\paragraph{B-Estimating the HR patches using the MMSE estimator}
For estimating the HR patch from a given LR patch $x_{LR}\in\R^{\tau^2}$, we first select the best component of the joint GGMM, such that the likelihood that $x_{LR}$ belongs to the $k*$-th component is maximal, i.e.
\begin{equation}
    k^*=\argmax_{k=1,...,K} w_k f\bracket{x_{LR}\mid \mu_{L}^k,\Sigma_{L}^k,\beta^k}.
\end{equation}
 Now, each HR patch can be estimated by the MMSE method thanks to Theorem \ref{theorem E[X|Y]} in Section \ref{section_MMSE_GGMM} based on the parameters of the generalized Gaussian mixture $k^*_i$
\begin{equation}
    \hat{x}_{HR}=\mu_{H,k^*}+\Sigma_{HL,k^*} \Sigma_{L,k^*}^{-1} \left(x_{LR}-\mu_{L,k^*}\right).
\end{equation}
\paragraph{C-Reconstructing HR image from HR patches}Finally, we reconstruct the high-resolution image from all estimated HR patches from the previous step. Let $x_{HR}=(x_{k,l})_{k,l=1}^{q\tau}\in\R^{q\tau,q\tau}$ be a two-dimensional high-resolution patch.
Then, we assign to each pixel $x_{k,l}$ the weight
$$
\rho_{k,l}\coloneqq\exp\Big(-\tfrac\gamma2\big((k-\tfrac{q\tau+1}{2})^2+(l-\tfrac{q\tau+1}{2})^2\big)\Big).
$$
After that, we add up for each pixel in the high resolution image the corresponding weighted pixel values and normalize the result by dividing by the sum of the weights.
\section{MMSE estimator with GGD} \label{section_MMSE_GGMM}
This section discusses a method to estimate the HR patches for step B in Section \ref{sec: SR}. 
Assume that the estimator $T\colon \R^d\to \R^D$ satisfies ${X}_{H} = T(X_{L})$ and $X=\bracket{X_H,X_L}$ has a GGD, which is selected as the best component of GGMM. Estimation $\hat{X}_{H}$ using the MMSE estimator is 
\begin{align}
    T_{MMSE}&\in \argmin_{T} \mathbb{E} \|X_{H}-T\bracket{X_{L}}\| _2^2. \label{mmse_1}
\end{align}
The Lehmann-Scheff\'e theorem~\cite{LS50} states that the general solution of the minimization problem \eqref{mmse_1} is given by $$T_{\text{MMSE}}= \mathbb{E}(X_{H}|X_{L}). $$ Since a generalized Gaussian distribution is also an elliptical distribution, the following theorem about the MMSE estimator $T_{MMSE}$ for GGD is a consequence of Theorem 8 in \cite{elliptical}.
\begin{theo}\label{theorem E[X|Y]}
Assume that $X=(X_{H},X_{L})\colon\Omega\to\R^{p}$ has a generalized Gaussian distribution $P_\theta$ with parameters $\theta=(\mu,\Sigma,\beta)$, where
$$
\mu=\Big(\begin{array}{c}\mu_{H}\\\mu_{L}\end{array}\Big),\quad\Sigma=\Big(\begin{array}{cc}\Sigma_{H}&\Sigma_{HL}\\\Sigma_{HL}^T&\Sigma_{L}\end{array}\Big)
$$
Then, for each $P_{X_{L}}$-almost every $x_{LR}$, we have that the conditional distribution
$P_{X_{H}|X_{L}=x_{LR}}$ is given by the generalized Gaussian distribution $P_{\hat \theta}$, where the parameters $\hat \theta=(\hat\mu,\hat\Sigma,\hat \beta)$ are given by
$$
\hat\mu=\mu_{H}+\Sigma_{HL}\Sigma_L^{-1}(x_{LR}-\mu_L), \hat\Sigma=\Sigma_H-\Sigma_{HL}\Sigma_L^{-1}\Sigma_{HL}^T, \hat\beta=\beta. 
$$
\end{theo} 
The MMSE estimator for GGD can be written as:
\begin{equation}
    T_{MMSE}=\mu_H+\Sigma_{HL} \Sigma_{L}^{-1} \left(x_{LR}-\mu_L\right).
\end{equation}
\section{Experimental Results}  Experimental results are given both on synthetic data of basic images such as Gold-hill, Barbara, Camera-man, and our real material data:  Fontainebleau sandstone (FS) and SiC Diamonds, which were presented in \cite{Johannes}. The observed LR image is generated from ground truth images with $q=2$ by the operator $A$ that is exactly given and defined as in \cite{Johannes}. In the training step, 
we use the upper left quarter of the HR and LR images.  
 \begin{table}[htp]
\caption{\label{tab_results_2dMMSE}
PSNRs of the reconstructions using MMSE and EPLL approaches for GMM and GGMM.  
} \label{talbe:MMSE}
\begin{center}
\begin{tabular}{||c||c|c|c||c|c||}
\hline  
\hline  
&Hill&Camera&Barbara&FS&Sics\\
\hline
\hline  
MMSE-GMM&$31.60$&$32.75$ & ${25.27}$ &33.09&28.00\\
\hline 
MMSE-GGMM&$\textbf{31.70}$& ${32.86}$&   ${25.33}$&\textbf{33.35}& \textbf{28.08} \\ 
\hline
EPLL-GMM&$31.62$&${32.91}$  & $\textbf{25.39}$&31.83&26.04\\ 
\hline 
EPLL-GGMM  & $31.58$ & $\textbf{32.94}$& ${25.33}$&31.89&26.06 \\ 
\hline  
\hline  
\end{tabular}
\end{center}
\end{table} 

Table 1 gives the PSNR values for the MMSE and EPLL methods using GMM and GGMM with $K=100$ components. 
For basic images, the MMSE-GGMM method gives higher PSNR values than the MMSE-GMM model, and they are not significantly different from EPLL-GGMM while our method does not require knowledge of $A$. 

Reconstructions of the material images are shown in Figures 1 and 2. We observe that the MMSE-GGMM results are sharper and visually better than the ones of the EPLL method. The PSNR values show that our method outperforms the EPLL-GGMM in this setting. This proves that our method does not need to learn parameters but can still achieve better results for the material data.
\begin{figure} [!h]
    \centering
    \begin{subfigure}[t]{0.14\textwidth}
\centering
\includegraphics[width=\textwidth]{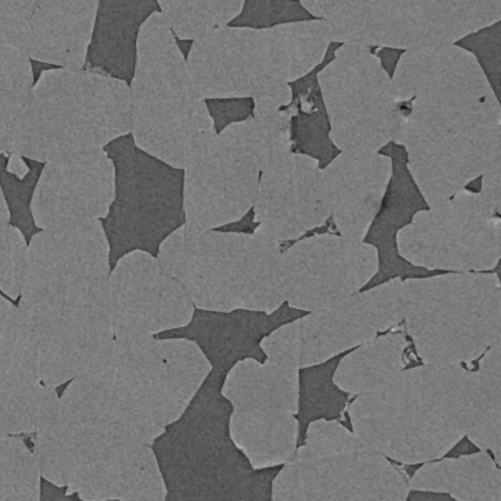} 
\subcaption{HR}
\end{subfigure}\hfill
    \begin{subfigure}[t]{0.14\textwidth}
\centering
\includegraphics[width=\textwidth]{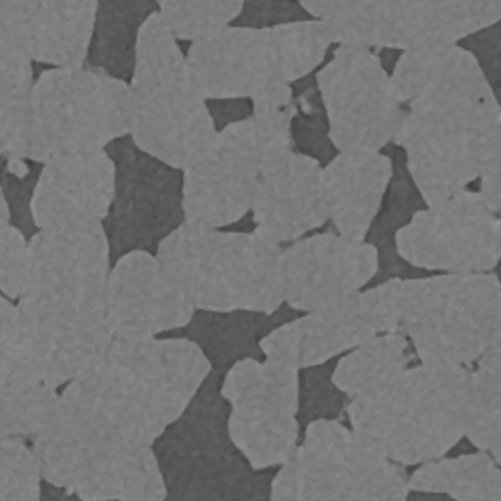} 
\subcaption{MMSE-GMM}
\end{subfigure}\hfill
    \begin{subfigure}[t]{0.14\textwidth}
\centering
\includegraphics[width=\textwidth]{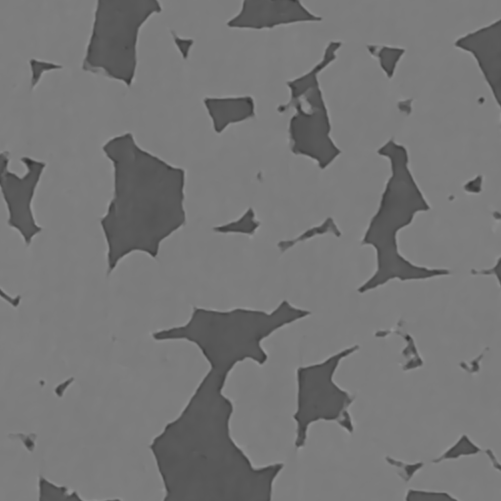}
\subcaption{MMSE-GGMM}
\end{subfigure}\\
    \begin{subfigure}[t]{0.14\textwidth}
\centering
\includegraphics[width=\textwidth]{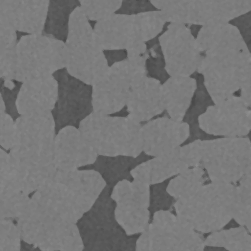} 
\subcaption{LR}
\end{subfigure}\hfill
    \begin{subfigure}[t]{0.14\textwidth}
\centering
\includegraphics[width=\textwidth]{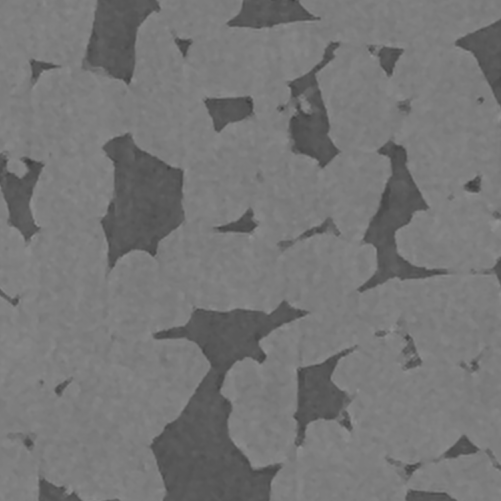} 
\subcaption{EPLL-GMM}
\end{subfigure}\hfill
    \begin{subfigure}[t]{0.14\textwidth}
\centering
\includegraphics[width=\textwidth]{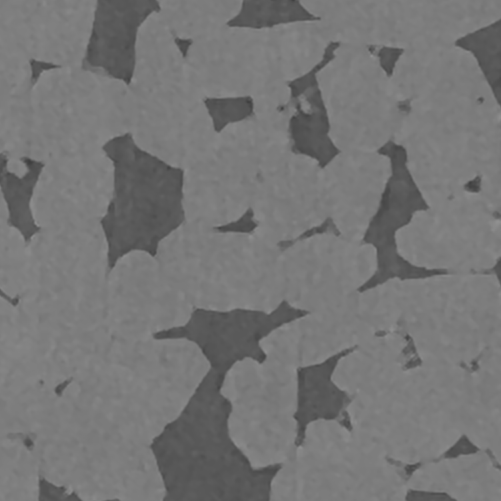}
\subcaption{EPLL-GGMM}
\end{subfigure}\\
    \caption{Reconstructions of 2D low resolution FS images by using MMSE and EPLL method.}
    \label{fig:FS MMSE}
\end{figure}  
\begin{figure} [!h]
    \centering
    \begin{subfigure}[t]{0.14\textwidth}
\centering
\includegraphics[width=\textwidth]{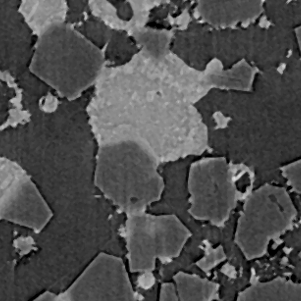} 
\subcaption{HR}
\end{subfigure}\hfill
    \begin{subfigure}[t]{0.14\textwidth}
\centering
\includegraphics[width=\textwidth]{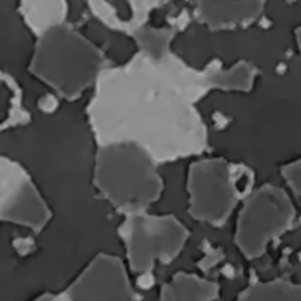} 
\subcaption{MMSE-GMM}
\end{subfigure}\hfill
    \begin{subfigure}[t]{0.14\textwidth}
\centering
\includegraphics[width=\textwidth]{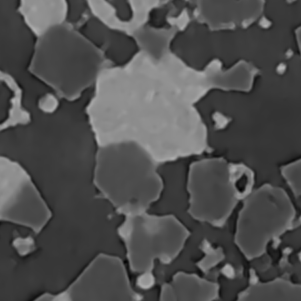}
\subcaption{MMSE-GGMM}
\end{subfigure}\\
    \begin{subfigure}[t]{0.14\textwidth}
\centering
\includegraphics[width=\textwidth]{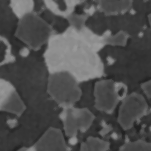} 
\subcaption{LR}
\end{subfigure}\hfill
    \begin{subfigure}[t]{0.14\textwidth}
\centering
\includegraphics[width=\textwidth]{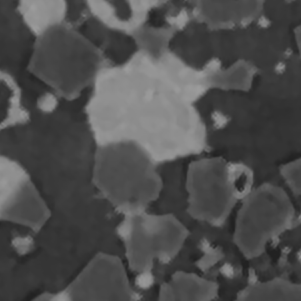} 
\subcaption{EPLL-GMM}
\end{subfigure}\hfill
    \begin{subfigure}[t]{0.14\textwidth}
\centering
\includegraphics[width=\textwidth]{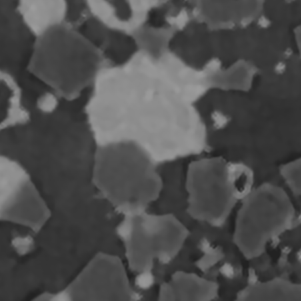}
\subcaption{EPLL-GGMM}
\end{subfigure}\\
    \caption{Reconstructions of 2D low resolution Sic Diamonds images by using MMSE and EPLL method.}
    \label{fig:FS MMSE}
\end{figure} 
 \section{Conclusion}
This paper proposed a new algorithm to perform image SR. We extended the image SR using the GMM method provided by Sandeep and Jacob \cite{Sandeep} to the GGMM model, which is learned by the FP-EM algorithm. Experiments on synthetic and material images demonstrate that our method is a promising solution for image SR. In future work, we will consider some deep learning approaches for super resolution with high magnification factor and high dimensional data.
\section*{Acknowledgment} Funding by the French Agence Nationale de la Recherche (ANR) under reference 
ANR-18-CE92-0050 SUPREMATIM is gratefully acknowledged.

\end{document}